\documentclass[aps,plb,preprintnumbers,nofootinbib]{revtex4}

\usepackage{amsmath,amsfonts,amssymb,bm}
\usepackage{graphicx}
\usepackage{color}
\usepackage{subfigure}
\usepackage{textcomp}
\usepackage{slashed}
\usepackage{enumitem}

\definecolor{purple}{rgb}{0.5,0,0.5}
\definecolor{blue}{rgb}{0.0,0,0.9}

\usepackage[colorlinks=true, pdfstartview=FitV, linkcolor=purple, citecolor= purple, urlcolor=blue]{hyperref}
\renewcommand{\baselinestretch}{1.5}

\begin{document}
	\title{\Large Study the decays of $\chi _{cJ}(J=0,1,2)$ to light meson pairs \\with SU(3) flavor symmetry/breaking analysis}
	\author{Bo Lan$^1$,~~Qin-Ze Song$^1$,~~Jin-Huan Sheng$^2$,~~Yi Qiao$^3$, and Ru-Min Wang$^{1,\dag}$\\
{\scriptsize $^1$College of Physics and Communication Electronics, Jiangxi Normal University, Nanchang, Jiangxi 330022, China}\\
 {\scriptsize $^2$School of Physics and Engineering, Henan University of Science and Technology, Luoyang, Henan 471000, China}\\
 {\scriptsize $^3$College of Physics and Electronic Information, Nanchang Normal University, Nanchang, Jiangxi 330032, China}\\
		$^\dag${\scriptsize Corresponding author.~~Email:ruminwang@sina.com}
}
	
\begin{abstract}
Based on available experimental results on $\chi _{cJ}(J=0,1,2)$ decays, we investigate the $\chi_{cJ}\to PP$, $VV$, $PV$, and $PT$  decays  by using SU(3) flavor symmetry/breaking approach, where $P$, $V$, and $T$ denote light pseudoscalar, vector, and tensor mesons, respectively.  With the decay amplitude relations determined by SU(3) flavor symmetry/breaking, we present the branching ratios for all $\chi_{cJ}\to PP$ and $\chi_{cJ}\to VV$ modes, including ones  without experimental data. While theoretical considerations strongly suppress or even forbid most $\chi_{cJ}\to PV$ and $PT$ decays, we also provide quantitative predictions constrained by existing experimental data. Our results are expected to be accessible in future experiments at BESIII and the planned Super Tau-Charm Facility.\\

\end{abstract}

\maketitle
	
\section{Introduction}
	The triplet state $\chi_{cJ} (J=0,1,2)$, as the lowest energy $P$-wave charmonium, has not been widely studied in the past because it cannot be directly obtained from the $e^{+}e^{-}$ collisions. However, $\psi(3686)$ provides a clean environment for the production of $\chi _{cJ}$ mesons via electromagnetic decays such as $\psi(3686)\rightarrow  {\gamma}{\chi_{cJ}}$.  As the $\psi(3686)$ has been widely studied and collected \cite{BES:2002psd,Ablikim:2012pj,Ablikim:2013ntc,BESIII:2017tvm,BESIII:2024lks}, $\chi_{cJ}$ decays has obtained more and better measurements, which has led to increased interest in the study of hadronic decays of the $\chi_{cJ}$ states \cite{BES:2005ukn,Luchinsky:2005bf,CLEO:2007rbf,CLEO:2008jma,BESIII:2010ank,Chen:2013gka,Gross:2022hyw}.

	The hadronic $\chi_{cJ}$ decays have great significance for studying strong force dynamics. On the one hand, the scale of charm quark mass ($\sim$1.5 GeV) is between the perturbative and non-perturbative QCD in theoretical calculations, it is not large enough for significant heavy quark expansion and not small enough for perturbation theory \cite{Cheng:2010ry,Cheng:2012wr}. On the other hand, most of the hadronic $\chi_{cJ}$ decays are suppressed by the Okubo-Zweig-Iizuka (OZI) rule \cite{Iizuka:1966fk}. Moreover, there are still theoretical and experimental discussions regarding the doubly OZI-suppressed decay and singly OZI-suppressed decay mechanisms in some processes \cite{Zhao:2007ze,Thomas:2007uy,Chen:2009ah,BESIII:2018pbx,BESII:2011hcd}. Due to these characteristics, understanding hadronic $\chi_ {cJ}$ decay mechanisms are important for improving theoretical models of both perturbative and non-perturbative QCD, and offer a valuable framework for testing phenomenological models and constraining theoretical parameters.

	The $\chi_{cJ}$ mesons decay into the two-meson states are particularly useful and they are relatively straightforward to detect and model theoretically. One important feature of charmonium hadronic decays into light mesons is that they are processes rich in gluons. The initial $c$ and $\bar{c}$ quark will annihilate into gluons, which then hadronize to produce the final state of light quarks \cite{Wang:2012qa}. Furthermore, the quantum numbers ($I^{G}J^{PC}$) of mesons produced by two photons and decaying into $PP,VV$ are restricted, they should be $0^{+}$ for isospin and $even^{++}$ for parity and charge conjugation, such as $\chi_{cJ}(J=even)$ mesons. Specifically, due to spin parity conservation, the $\chi_{c1}$ cannot decay into two pseudoscalar mesons. And they violate the so-called helicity selection rule, which is defined as $\sigma=(-1)^{J}{P}$ \cite{Chernyak:1981zz}, and require $\sigma^{initial}=\sigma_{1}\sigma_{2}$, where $J$ and $P$ are respectively the spin and parity of the particle.

	The theoretical calculations of charmonium decays have always been quite difficult, the non-relativistic QCD (NRQCD) framework \cite{Caswell:1985ui,Brambilla:1999xf} and phenomenological approaches are suitable and effective to study the charmonium decays. Previous studies of hadronic $\chi_{cJ}$ decays used various theoretical frameworks, including parametrization schemes \cite{Zhao:2005im,Zhao:2007ze}, intermediate hadronic loops \cite{Liu:2009vv,Chen:2009ah}, quark pair creation model with perturbative QCD (pQCD) framework \cite{Zhou:2004mw}, and a description within the effective field theory framework \cite{Kivel:2024iqe}, etc. In Refs. \cite{Zhao:2005im,Zhao:2007ze}, parametrization schemes were proposed to further understand the mechanism of violating OZI rules in the two body decays of $\chi_{cJ}\to PP,VV,SS$, where $S$ denotes scalar meson.  These studies suggest that, in addition to singly OZI-suppressed processes, doubly OZI-suppressed processes may play a significant role in the production of isospin-0 light meson pairs, such as $\chi_{c1}\rightarrow {f_{0}}{f'_{0}}$, $\omega\omega$, $\phi\phi$, $\omega\phi$, $\eta\eta$, $\eta\eta'$ and $\eta'\eta'$.
	
In this paper, we perform a SU(3) flavor symmetry and breaking analysis of charmonium decay processes $\chi_{cJ} \to PP$, $VV$, $PV$, and $PT$. In Sec. \ref{sec:PPVV}, the fundamental amplitude relations and theoretical frameworks governing the  $\chi_{cJ} \to PP,VV$  decays are given, and then the numerical results of the $\chi_{cJ} \to PP$, $VV$ decays are listed from present experimental data.
In Sec. \ref{sec:PVPT},  corresponding amplitude relations and numerical results of the $\chi_{cJ} \to PV$, $PT$ decays are given. The final conclusions can be found in Sec.~\ref{sec:Conclusion}.

\section{ The $\chi_{cJ} \to PP,VV$ Decays  } \label{sec:PPVV}
\subsection{Amplitude Relations of $\chi_{cJ} \to PP,VV$}   \label{subsec:PPVVAR}
	The pseudoscalar and vector meson octet states under the SU(3) flavor symmetry of $u,d,s$ quarks
	can be written as
	\begin{eqnarray}
		P &=&\left(
		\begin{array}{ccc}
			\frac{\pi ^{0}}{\sqrt{2}}+\frac{\eta _{8}}{\sqrt{6}}+\frac{\eta _{1}}{\sqrt{3%
			}} & \pi ^{+} & K^{+} \\
			\pi ^{-} & -\frac{\pi ^{0}}{\sqrt{2}}+\frac{\eta _{8}}{\sqrt{6}}+\frac{\eta
				_{1}}{\sqrt{3}} & K^{0} \\
			K^{-} & \bar{K}^{0} & -\frac{2\eta _{8}}{\sqrt{6}}+\frac{\eta _{1}}{\sqrt{3}}%
		\end{array}
		\right) , \label{P}
		\end{eqnarray}
		and
 	\begin{eqnarray}
		V &=&\left(
		\begin{array}{ccc}
			\frac{\rho ^{0}}{\sqrt{2}}+\frac{\omega _{8}}{\sqrt{6}}+\frac{\omega _{1}}{%
				\sqrt{3}} & \rho ^{+} & K^{\ast +} \\
			\rho ^{-} & -\frac{\rho ^{0}}{\sqrt{2}}+\frac{\omega _{8}}{\sqrt{6}}+\frac{%
				\omega _{1}}{\sqrt{3}} & K^{\ast 0} \\
			K^{\ast -} & \bar{K}^{\ast 0} & -\frac{2\omega _{8}}{\sqrt{6}}+\frac{\omega
				_{1}}{\sqrt{3}}
		\end{array}
		\right), \label{V}
		\end{eqnarray}	
where $\eta_{1}$-$\eta_{8}$ and $\omega_{1}$-$\omega_{8}$ denote the unmixed states, and according to Particle Data Group (PDG) \cite{ParticleDataGroup:2024cfk}, the mixing of $\eta$-$\eta'$ and $\phi$-$\omega$ are described as follows.
\begin{eqnarray}
	\eta  &=&\left( \eta _{8}\cos \theta _{P}-\allowbreak \eta _{1}\sin \theta
	_{P}\right),  \\
	\eta' &=&\left( \eta _{8}\sin \theta _{P}+\allowbreak \eta
	_{1}\cos \theta _{P}\right),
\end{eqnarray}
and
\begin{eqnarray}
	\phi  &=&\left( \omega _{8}\cos \theta _{V}-\allowbreak \omega _{1}\sin \theta_{V}\right),  \\
	\omega &=&\left( \omega _{8}\sin \theta _{V}+\allowbreak \omega_{1}\cos \theta _{V}\right).
\end{eqnarray}
 Many theoretical works consider the mesons $\omega$ and $\eta$ as pure octet components, experimental observations show that $\omega$ and $\eta$ are actually mixtures of both octet and singlet components, with a specific mixing angle. Therefore, we discuss the branching ratio results that incorporate the $\eta$-$\eta'$ and $\phi$-$\omega$ mixing with the mixing angle values from the PDG, $\theta _{P}=[-20^\circ,-10^\circ]$ and $\theta _{V}=36.5^\circ$  \cite{ParticleDataGroup:2024cfk}.

\begin{figure}[hb]
	\centering
	\includegraphics[width=0.75\linewidth]{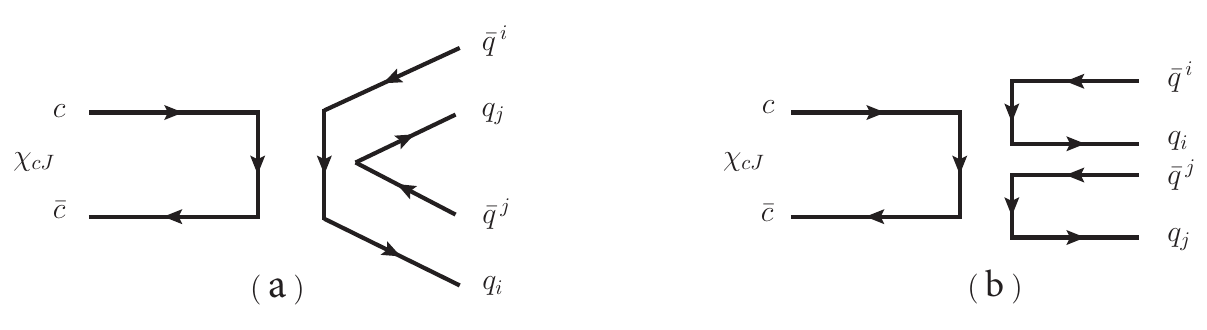}
	\caption{ The singly OZI disconnected  (a) and  doubly OZI disconnected
 (b) diagrams of the $\chi _{cJ}\rightarrow MM$ decays. }
	\label{fig:1}
\end{figure}

As shown in Fig. \ref{fig:1},  the singly OZI disconnected and  doubly OZI disconnected  decay modes are considered.The Fig. \ref{fig:1}(a) shows a typical singly OZI process, where gluons produced by initial state annihilation create new quark pairs and undergo quark exchange in the final state. In Fig. \ref{fig:1}(b), two quark pairs are created by gluons, and recoil without quark exchanges, which show a doubly OZI process. Based on the SU(3) flavor symmetry, the decay  amplitudes of the decay $\chi _{cJ}\rightarrow  MM$ can be parameterized as
\begin{eqnarray}	
		\mathcal{A}^{s}(\chi _{cJ}\rightarrow  M_1M_2)=a^{M}_{1J}M_{1j}^{i}M_{2i}^{j}+a^{M}_{2J}M_{1i}^{i}M_{2j}^{j},
\end{eqnarray}
 where $M=P/V$,  $i,j=1,2,3$ correspond to the matrix elements in Eqs.(\ref{P}-\ref{V}).The $a^{M}_{1J}$ and  $a^{M}_{2J}$ are non-perturbative coefficients corresponding to the decay modes in Fig. \ref{fig:1}(a) and Fig. \ref{fig:1}(b), respectively. The specific decay amplitudes for $\chi _{cJ}\rightarrow  {P}{P},{V}{V}$ with the SU(3) flavor symmetry are listed in the second column of Tab. \ref{Tab:AmpPPVV}.

\begin{table}[hbt]
		\renewcommand\arraystretch{1.5}
		\tabcolsep 0.25in
		\caption{The SU(3) symmetry and breaking amplitudes of $\chi_{cJ}\to P_1P_2$ and $\chi_{cJ}\to V_1V_2$.}
		\begin{center}
		\resizebox{\textwidth}{!}{
			\begin{tabular}{lll}
				\hline\hline
				Decay modes                                                      & Symmetry amplitudes                                                    & Breaking amplitudes                                                     \\ \hline
				$\chi _{cJ}\rightarrow  \pi^{+}\pi^{-}/\rho^{+}\rho^{-}$         & $2a^{M}_{1J}$                                                          & $2b^{M}_{J}$                                                            \\
				$\chi _{cJ}\rightarrow  K^{+}K^{-}/K^{*+}K^{*-}$                 & $2a^{M}_{1J}$                                                          & $-b^{M}_{J}$                                                            \\
				$\chi _{cJ}\rightarrow  K^{0}\bar{K}^{0}/K^{*0}\bar{K}^{*0}$     & $2a^{M}_{1J}$                                                          & $-b^{M}_{J}$                                                            \\
				$\chi _{cJ}\rightarrow  \pi^{0}\pi^{0}/\rho^{0}\rho^{0}$         & $\sqrt{2}a^{M}_{1J}$                                                   & $\sqrt{2}b^{M}_{J}$                                                     \\
				$\chi _{cJ}\rightarrow  \eta_{1} \eta_{1}/\omega_{1} \omega_{1}$ & $\sqrt{2}(a^{M}_{1J}+3a^{M}_{2J})$                                     & $0$                                                                     \\
				$\chi _{cJ}\rightarrow  \eta_{1} \eta_{8}/\omega_{1} \omega_{8}$ & $0$                                                                    & $2\sqrt{2}b^{M}_{J}$                                                    \\
				$\chi _{cJ}\rightarrow  \eta_{8} \eta_{8}/\omega_{8} \omega_{8}$ & $\sqrt{2}a^{M}_{1J}$                                                   & $-\sqrt{2}b^{M}_{J}$                                                    \\ \hline
				$\chi _{cJ}\rightarrow  \eta \eta/\phi \phi$                     & $\sqrt{2}a^{M}_{1J}+\frac{3\sqrt{2}}{2}a^{M}_{2J}(1-\cos2\theta _{M})$ & $-\sqrt{2}b^{M}_{J}\cos\theta _{M}(4\sin \theta _{M}+\cos \theta _{M})$ \\
				$\chi _{cJ}\rightarrow  \eta \eta'/\phi \omega$                  & $-\frac{3\sqrt{2}}{2}a^{M}_{2J}\sin2\theta _{M}$                       & $\sqrt{2}b^{M}_{J}(2\cos 2\theta _{M}-\frac{1}{2} \sin 2\theta _{M})$   \\
				$\chi _{cJ}\rightarrow  \eta' \eta'/\omega \omega$               & $\sqrt{2}a^{M}_{1J}+\frac{3\sqrt{2}}{2}a^{M}_{2J}(1+\cos2\theta_{M})$  & $\sqrt{2}b^{M}_{J}\sin \theta _{M}(4\cos\theta _{M}-\sin \theta _{M})$  \\ \hline
			\end{tabular}}
			\end{center}\label{Tab:AmpPPVV}
	\end{table}

	The SU(3) flavor symmetry assumes that $u$, $d$, and $s$ quarks have equal masses. SU(3) flavor symmetry breaking is given by  the current
quark mass term in the QCD Lagrangian with  $m_{u,d}\ll m_{s}$ in  the Standard Model. In order to achieve a more accurate analysis, we take into account the symmetry breaking.  The diagonalized mass matrix can be expressed as \cite{Gronau:1995hm,Xu:2013dta,He:2014xha}
		\begin{equation}
		\left(
		\begin{array}{ccc}
			m_{u} & 0 & 0  \\
			0 & m_{d} & 0  \\
			0 & 0 & m_{s}
		\end{array}
		\right)=\frac{1}{3}(m_{u}+m_{d}+m_{s})I + \frac{1}{2}(m_{u}-m_{d})X + \frac{1}{6}(m_{u}+m_{d}-2m_{s})W,
	\end{equation}
	where $I$ is the identity matrix, and $X$ and $W$ represented as
	\begin{equation}
		X=\left(
		\begin{array}{ccc}
			1 & 0 & 0  \\
			0 & -1 & 0  \\
			0 & 0 & 0
		\end{array}
		\right),~~~~~~
	W=\left(
		\begin{array}{ccc}
			1 & 0 & 0  \\
			0 & 1 & 0  \\
			0 & 0 & -2
		\end{array}
	\right).\label{Eq:XW}
	\end{equation}
	Considering the mass difference between the $s$ quark and the $u$, $d$ quarks, the $W$ part of the diagonalized mass matrix to account for symmetry breaking is included. Thus, the amplitude of the breaking part can be written as
\begin{eqnarray}	
	\mathcal{A}^{b}(\chi _{cJ}\rightarrow  M_1M_2)=b^{M}_{J}M_{1j}^{i}W_{k}^{j}M_{2i}^{k},
\end{eqnarray}
where $b^{M}_{J}$ is the non-perturbative coefficient of the breaking term. The decay amplitudes for $\chi _{cJ}\rightarrow  {P_1}{P_2},{V_1}{V_2}$ with SU(3) breaking are given in the third column of Tab. \ref{Tab:AmpPPVV},  and it is worth noting that the minus signs of the breaking terms are from  the mathematical structure of the $W$ matrix given in Eq. (\ref{Eq:XW}). Furthermore, we follow Ref. \cite{Gronau:1994rj,Wang:2020gmn} to make a similar convention regarding the final state being identical particles, where the amplitude of the final state of identical particles is multiplied by a coefficient with $\sqrt{2}$.

In general,  the involved  three parameters ($a^{M}_{1J}$, $a^{M}_{2J}$, $b^{M}_{J}$) are complex. Since an overall phase
can be removed without losing generality, we set $a^{M}_{1J}$ to be real, then there are five real
independent parameters are considered in data analysis
\begin{eqnarray}	
	a^{M}_{1J},~~ a^{M}_{2J}e^{i\alpha_{JM}}, ~~b^{M}_{J}e^{i\beta_{JM}}.
\end{eqnarray}
Where $\alpha_{JM}$ is the relative phase between $a^{M}_{1J}$ and $a^{M}_{2J}$, and $\beta_{JM}$ is the relative phase between $a^{M}_{1J}$ and $b^{M}_{J}$. It should be noted that when only one decay mode is contributed, the relative phase does not appear, and its own phase disappears in the form of modulus in the branching ratio calculations, so there is no need for special definition. The numerical results of branching ratio are provided by Monte Carlo simulation method, and all experimental inputs are from PDG \cite{ParticleDataGroup:2024cfk}. Each process inputs a large number of random samples to ensure that the distribution of numerical results consists of at least 10000 sets of valid results. More detailed discussions will be provided in the following contents.
	The branching ratio is calculated using the fundamental two-body decay formula, which can be expressed as
   \begin{equation}
   	\mathcal{B}(\chi _{cJ}\to M_1M_2)=\frac{\left\vert \vec{p}\right\vert }{8\pi M_{\chi
   			_{cJ}}^{2}\Gamma _{\chi _{cJ}}}\Big| \mathcal{A}(\chi _{cJ}\to M_1M_2)\Big| ^{2},  
   \end{equation}
   where $\Gamma _{\chi_{cJ}}$ is the decay width of the $\chi_{cJ}$\ meson, and the center-of-mass momentum $|\vec{p}|\equiv\frac{\sqrt{[M_{\chi _{cJ}}^{2}-(m_{M_1}+m_{M_2})^{2}][M_{\chi_{cJ}}^{2}-(m_{M_1}-m_{M_2})^{2}]}}{2M_{\chi _{cJ}}}$.

 \subsection{Numerical results of the $\chi_{cJ}\to PP$ decays}  \label{subsec:PP}

	Numerical results of the $\chi _{cJ}\rightarrow  {P}{P}$ decays will be given in this section. The experimental branching ratios are provided in the second column in Tab. \ref{Tab:data0PP} for $\chi _{c0}$ decays and Tab. \ref{Tab:data2PP} for $\chi _{c2}$ decays from PDG \cite{ParticleDataGroup:2024cfk}. Base on SU(3) symmetry and breaking amplitudes, we have obtained and analyzed two sets of branching ratio results for $\chi _{cJ}\rightarrow  {P}{P}$ decays. The results considering SU(3) symmetry are shown in the third column of Tab. \ref{Tab:data0PP} and Tab. \ref{Tab:data2PP}, while those incorporating breaking effects are listed in the fourth columns.  Furthermore, the decays of $\chi_{c1}$ ($J^{PC}=1^{++}$) into two pseudoscalar mesons is forbidden by spin-parity conservation and helicity selection rule, therefore, they are not considered in this analysis.

\begin{table}[h]
		\caption{The branching ratios for the $\chi_{c0} \to PP$ decays within  1$\sigma$ error (in units of $10^{-3}$). $^\sharp$indicates that experimental data was not used to derive the numerical results.}
		\centering%
		\begin{center}
		\renewcommand\arraystretch{0.99}\tabcolsep 0.15in
			\begin{tabular}{lccc}
				\hline\hline
				                                             & Exp. data \cite{ParticleDataGroup:2024cfk} &         Symmetry          &     Including breaking     \\ \hline
				$\mathcal{B}(\chi_{c0}\to \pi^{+}\pi^{-})$   &                  $\cdots$                  &      $6.33\pm 0.33$       &       $5.67\pm0.27$        \\
				$\mathcal{B}(\chi_{c0}\to \pi^{0}\pi^{0})$   &                  $\cdots$                  &      $3.17\pm 0.16$       &       $2.83\pm0.13$        \\
				$\mathcal{B}(\chi_{c0}\to \pi\pi)$           &               $8.50\pm 0.40$               &   $9.50\pm 0.49^\sharp$   &       $8.50\pm0.40$        \\
				$\mathcal{B}(\chi_{c0}\to K^{+}K^{-})$       &               $6.07\pm 0.33$               &      $6.08\pm 0.32$       &       $6.07\pm0.33$        \\
				$\mathcal{B}(\chi_{c0}\to K^{0}\bar{K}^{0})$ &                  $\cdots$                  &      $6.08\pm 0.32$       &       $6.07\pm0.33$        \\
				$\mathcal{B}(\chi_{c0}\to K\bar{K})$         &                  $\cdots$                  &      $12.16\pm 0.63$      &       $12.14\pm0.66$       \\
				$\mathcal{B}(\chi_{c0}\to \eta \eta)$        &               $3.01\pm 0.25$               &      $2.91\pm 0.15$       &       $2.92\pm0.16$        \\
				$\mathcal{B}(\chi_{c0}\to \eta \eta')$       &               $0.09\pm 0.01$               &      $0.09\pm 0.01$       &       $0.09\pm0.01$        \\
				$\mathcal{B}(\chi_{c0}\to \eta' \eta')$      &               $2.17\pm 0.12$               &      $2.17\pm 0.12$       &       $2.17\pm0.12$        \\ \hline
				$a^{P}_{10}$ $(10^{-2})$                     &                  $\cdots$                  &      $5.41\pm 0.28$       &       $5.30\pm 0.27$       \\
				$a^{P}_{20}$ $(10^{-2})$                     &                  $\cdots$                  &      $1.48\pm 0.56$       &       $1.12\pm 1.04$       \\
				$b^{P}_{0}$ $(10^{-2})$                      &                  $\cdots$                  &         $\cdots$          &       $0.40\pm 0.35$       \\
				$\lvert\alpha _{0P}\rvert$                   &                  $\cdots$                  & $(119.75 \pm 9.74)^\circ$ & $(142.09 \pm 37.82)^\circ$ \\
				$\lvert\beta _{0P}\rvert$                    &                  $\cdots$                  &         $\cdots$          & $(136.36 \pm 43.54)^\circ$ \\ \hline
			\end{tabular}
		\end{center}\label{Tab:data0PP}
%
		\caption{The branching ratios for the $\chi_{c2} \to PP$ decays within  1$\sigma$ error (in units of $10^{-4}$). $^\sharp$indicates that experimental data was not used to derive the numerical results.}
		\centering
			\begin{center}
		\renewcommand\arraystretch{0.98}\tabcolsep 0.15in
			\begin{tabular}{lccc}
				\hline\hline
				                                              & Exp. data \cite{ParticleDataGroup:2024cfk} &         Symmetry          &     Including breaking     \\ \hline
				$\mathcal{B}(\chi_{c2}\to \pi^{+}\pi^{-})$    &                  $\cdots$                  &      $11.67\pm 0.47$      &      $15.13\pm 0.67$       \\
				$\mathcal{B}(\chi_{c2}\to \pi^{0}\pi^{0})$    &                  $\cdots$                  &      $5.83\pm 0.24$       &       $7.57\pm 0.33$       \\
				$\mathcal{B}(\chi_{c2}\to \pi\pi)$            &              $22.70\pm 1.00$               &  $17.50\pm 0.71^\sharp$   &      $22.70\pm 1.00$       \\
				$\mathcal{B}(\chi_{c2}\to K^{+}K^{-})$        &              $10.20\pm 1.50$               &      $11.24\pm 0.46$      &      $10.20\pm 1.50$       \\
				$\mathcal{B}(\chi_{c2}\to  K^{0}\bar{K}^{0})$ &                  $\cdots$                  &      $11.24\pm 0.46$      &      $10.20\pm 1.50$       \\
				$\mathcal{B}(\chi_{c2}\to  K\bar{K})$         &                  $\cdots$                  &      $22.48\pm 0.91$      &       $20.39\pm3.00$       \\
				$\mathcal{B}(\chi_{c2}\to \eta \eta)$         &               $5.50\pm 0.50$               &      $5.22\pm 0.22$       &       $5.42\pm 0.42$       \\
				$\mathcal{B}(\chi_{c2}\to \eta \eta')$        &               $0.22\pm 0.05$               &      $0.22\pm 0.05$       &       $0.22\pm 0.05$       \\
				$\mathcal{B}(\chi_{c2}\to \eta' \eta')$       &               $0.46\pm 0.06$               &      $0.46\pm 0.06$       &       $0.46\pm 0.06$       \\ \hline
				$a^{P}_{12}$ $(10^{-2})$                      &                  $\cdots$                  &      $1.02\pm 0.04$       &       $1.03\pm 0.08$       \\
				$a^{P}_{22}$ $(10^{-2})$                      &                  $\cdots$                  &      $0.35\pm 0.11$       &       $0.32\pm 0.16$       \\
				$b^{P}_{2}$ $(10^{-2}) $                      &                  $\cdots$                  &         $\cdots$          &       $0.14\pm 0.08$       \\
				$\lvert\alpha _{2P}\rvert$                    &                  $\cdots$                  & $(170.17 \pm 9.74)^\circ$ & $(165.58 \pm 14.32)^\circ$ \\
				$\lvert\beta _{2P}\rvert$                     &                  $\cdots$                  &         $\cdots$          & $(31.51 \pm 30.94)^\circ$  \\ \hline
			\end{tabular}
		\end{center}\label{Tab:data2PP}
	\end{table}

Firstly, analyze the results within the SU(3) flavor symmetry.  The  experimental data of the $\chi _{c0}\to K^+K^-,$ $\eta\eta,\eta\eta',\eta'\eta'$ and $\chi _{c2}\to K^+K^-,\eta\eta,\eta\eta',\eta'\eta'$ decays can be explained within $1\sigma$ error. After using the constrained non-perturbative parameters from  the data of $\chi _{c0}\to K^+K^-,\eta\eta,\eta\eta',\eta'\eta'$ or $\chi _{c2}\to K^+K^-,\eta\eta,\eta\eta',\eta'\eta'$ decays,  the predicted branching ratios of the  $\chi _{c0}\to \pi\pi$ and $\chi _{c2}\to \pi\pi$ decays  slightly deviate their experimental data within $1\sigma$ error, nevertheless, they could be consistent within about $2\sigma$ or $3\sigma$ error bars.
All predicted branching ratios also falling within a relatively accurate range. This suggests that the SU(3) flavor symmetry exhibits good conformity with the overall decay modes.  In particular, the $a^{P}_{1J}$ term exhibits dominance across most decay channels. The maximum value of the ratio $\frac{a^{P}_{2J}}{a^{P}_{1J}}$ reach 37\% for the $\chi _{c0}$ decays and 45\% for the $\chi _{c2}$ decays, and they are supported by suppressed from doubly OZI \cite{BESIII:2017rqn,Zheng:2017ffd}. Within SU(3) flavor symmetry, amplitude form of $\eta \eta'$ only has $a^{P}_{2J}$ component, which also conforms to the result of suppression.

After considering  the SU(3) flavor breaking effects, as listed in the fourth columns of  Tab. \ref{Tab:data0PP} and Tab. \ref{Tab:data2PP},   all experimental data of  both $\chi _{c0}$ and $\chi _{c2}$ decays including the $\pi\pi$ modes, can be explained within $1\sigma$ error bar.  In terms of the ratio  $\frac{b^{M}_{J}}{a^{M}_{1J}}$ as the breaking ratio, the maximum value is 14\% in the $\chi _{c0}$  decays and 22\% in the $\chi _{c2}$  decays. Although $b^{M}_{J}$ term contributions are small compared to the $a^{P}_{2J}$, they provide better fits for the relative experimental data within $1\sigma$ error. As a result, there is an increased error range for certain channels, such as $\chi_{c2}\to K^{+}K^{-}$ and $\chi_{c2}\to  K^{0}\bar{K}^{0}$. It implies that while the breaking contributions are small, they should not be ignored.

\begin{figure}[t]
\begin{center}
\includegraphics[scale=0.2]{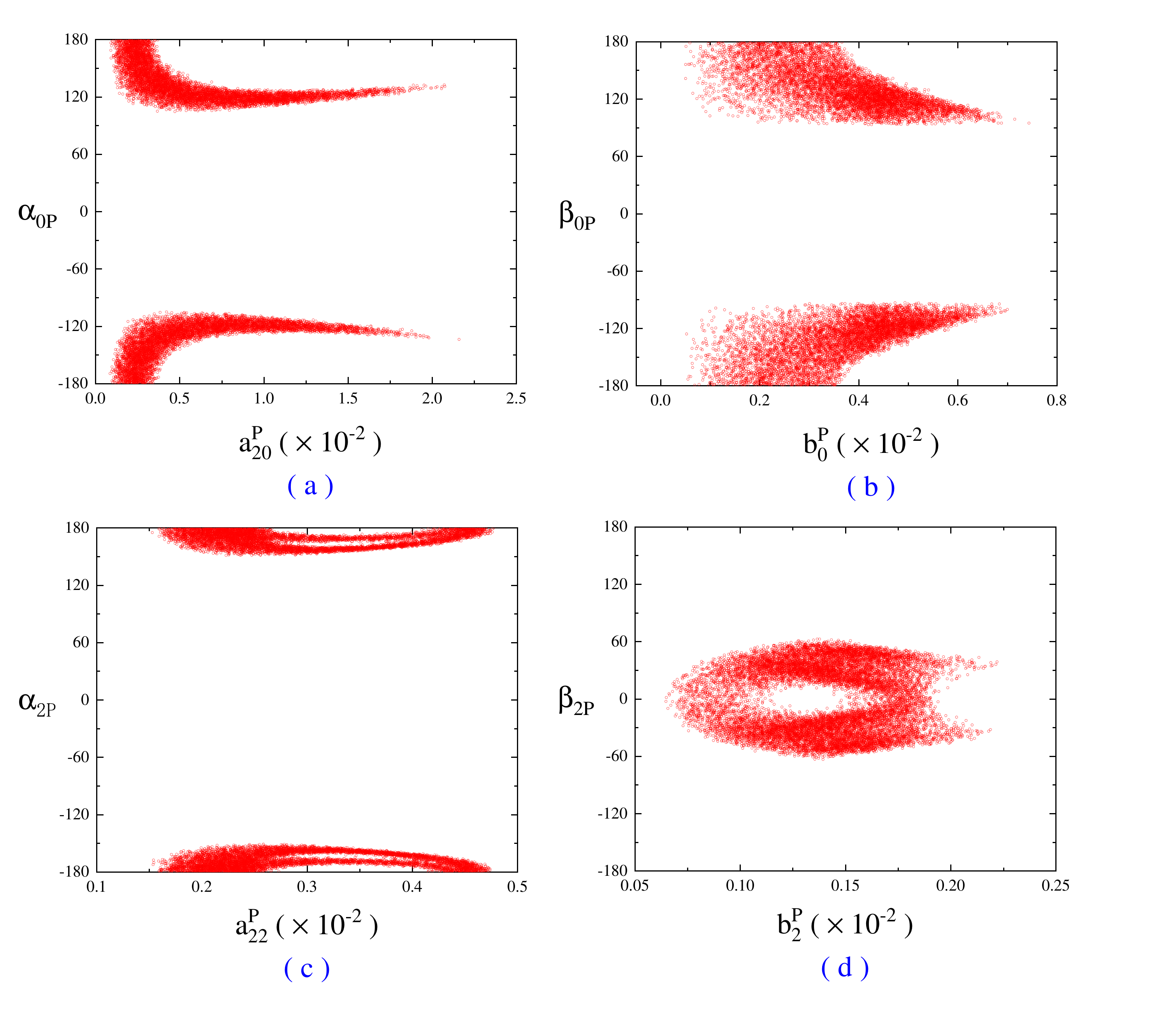}
\end{center}
\caption{The parameter correlations in the $\chi _{c0,2}\rightarrow PP$ decays.
 }\label{fig:PP}
\end{figure}

For the non-perturbative parameters  $a^{P}_{1J}$, $a^{P}_{2J}$, $b^{P}_{J}$, $\lvert\alpha _{JP}\rvert$ and $\rvert\beta _{JP}\rvert$,   their allowed ranges are listed in the last five lines of Tab. \ref{Tab:data0PP} and Tab. \ref{Tab:data2PP}. Please note that their allowed values   are interrelated. Their interrelations are shown  in Fig. \ref{fig:PP}.

Noted that the recent measurements on BESIII \cite{BESIII:2025tho} report precision branching ratios
\begin{align*}
\mathcal{B}(\chi_{c0}\to K^{+}K^{-})&=(6.36\pm0.15)\times10^{-3},~~~~~~~~~~
\mathcal{B}(\chi_{c0}\to \pi^{+}\pi^{-})=(6.06\pm0.15)\times10^{-3}, \\
\mathcal{B}(\chi_{c2}\to K^{+}K^{-})&=(1.22\pm0.03)\times10^{-3},~~~~~~~~~~
\mathcal{B}(\chi_{c2}\to \pi^{+}\pi^{-})=(1.61\pm0.04)\times10^{-3},
\end{align*}
which are not used to constrain the non-perturbative coefficients and to predict the not-yet-measured branching ratios in this work.  Our predictions of $\mathcal{B}(\chi_{c0,2}\to \pi^{+}\pi^{-})$ are  consistent with above data  within $1\sigma$ error bar. Above experimental data of $\mathcal{B}(\chi_{c0}\to K^{+}K^{-})$ or $\mathcal{B}(\chi_{c2}\to K^{+}K^{-})$ are consistent with ones from PDG \cite{ParticleDataGroup:2024cfk} within $1\sigma$ or $1.2\sigma$ error bar(s).

In previous studies, several results provide useful reference points. For instance, Ref. \cite{Zhao:2007ze} proposed parametrization schemes to further understand the mechanisms that violate the OZI rule. Their work deepened the understanding of charmonium decays mechanisms by defining the relative strength $r$ and some other physical quantities based on the SU(3) flavor approach. They provided good fit results based on experimental data at the time, and provide useful insights into the channels of the isospin-0 light meson pairs. With the update of experimental measurements, we have provided more theoretical predictions under the SU(3) flavor approach using the latest experimental data. Our predicted results are expected to contribute to future research on $\chi_{cJ}\to PP$ decays.

\subsection{Numerical results of $\chi_{cJ}\to VV$} \label{sec:VV}

For $\chi_{cJ}\to VV$ decays, they are similar to $PP$ channels, but with some key differences. The $\chi_{c1}\to VV$ decays  are suppressed compared to  $\chi_{c0,2}\to VV$ decays since they violate helicity selection rule \cite{Chernyak:1981zz}, much like their $PP$ channels. However, interestingly, $\chi_{c1}\to VV$ decays are not forbidden as initially expected. As experimental data has accumulated, increasing observations have revealed significant discrepancies between the data and the predictions based on the selection rule. One possible reason for the failure of the perturbative approach here could be that, although the charm quark is relatively heavy, it does not meet the mass threshold required by pQCD \cite{Liu:2009vv}. This suggests that there might be other mechanisms at play, such as higher-order contributions, final-state interactions, or long-distance effects, that could contribute to processes typically forbidden by the helicity selection rule. In any case, we present the branching ratio results for $\chi_{c0,1,2}\to VV$ based on the SU(3) flavor symmetry/breaking in this section.

The experimental data are given in the second columns of Tab. \ref{Tab:data0VV}, Tab. \ref{Tab:data1VV} and Tab. \ref{Tab:data2VV} for the $\chi _{c0}\to  VV$,  $\chi _{c1}\to  VV$ and  $\chi _{c2}\to  VV$ decays, respectively. Our branching ratio predictions based on SU(3) flavor symmetry are listed in the third columns of these tables, while predictions  including breaking effects are displayed in the fourth columns.
	\begin{table}[t]
		\caption{The branching ratios for the $\chi_{c0} \to VV$ decays within  1$\sigma$ error (in units of $10^{-3}$).}
		\centering
			\begin{center}
		\renewcommand\arraystretch{0.98}\tabcolsep 0.15in
			\begin{tabular}{lccc}
				\hline\hline
				                                                & Exp. data  \cite{ParticleDataGroup:2024cfk} &          Symmetry          &     Including breaking     \\ \hline
				$\mathcal{B}(\chi_{c0}\to \rho^{+}\rho^{-})$    &                  $\cdots$                   &       $2.04\pm 0.37$       &       $2.28\pm 1.56$       \\
				$\mathcal{B}(\chi_{c0}\to \rho^{0}\rho^{0})$    &                  $\cdots$                   &       $1.02\pm 0.18$       &       $1.14\pm 0.78$       \\
				$\mathcal{B}(\chi_{c0}\to \rho\rho)$            &                  $\cdots$                   &       $3.06\pm 0.55$       &       $3.42\pm 2.33$       \\
				$\mathcal{B}(\chi_{c0}\to K^{*+}K^{*-})$        &                  $\cdots$                   &       $1.95\pm 0.35$       &       $1.70\pm 0.60$       \\
				$\mathcal{B}(\chi_{c0}\to  K^{*0}\bar{K}^{*0})$ &               $1.70\pm 0.60$                &       $1.95\pm 0.35$       &       $1.70\pm 0.60$       \\
				$\mathcal{B}(\chi_{c0}\to  K^{*}\bar{K}^{*})$   &                  $\cdots$                   &       $3.90\pm 0.70$       &       $3.40\pm 1.20$       \\
				$\mathcal{B}(\chi_{c0}\to \phi \phi)$           &              $0.848\pm 0.031$               &       $0.85\pm 0.03$       &       $0.85\pm 0.03$       \\
				$\mathcal{B}(\chi_{c0}\to \phi \omega)$         &              $0.142\pm 0.013$               &      $0.142\pm 0.013$      &      $0.142\pm 0.013$      \\
				$\mathcal{B}(\chi_{c0}\to \omega \omega)$       &               $0.97\pm 0.11$                &       $0.97\pm 0.11$       &       $0.97\pm 0.11$       \\ \hline
				$a^{V}_{10}$ $(10^{-2})$                        &                  $\cdots$                   &       $3.24\pm 0.37$       &       $2.98\pm 0.85$       \\
				$a^{V}_{20}$ $(10^{-2})$                        &                  $\cdots$                   &       $0.86\pm 0.06$       &       $0.85\pm 0.14$       \\
				$b^{V}_{0}$ $(10^{-2})$                         &                  $\cdots$                   &          $\cdots$          &       $0.78\pm 0.78$       \\
				$\lvert\alpha _{0V}\rvert$                      &                  $\cdots$                   & $(106.57 \pm 16.04)^\circ$ & $(120.89 \pm 59.01)^\circ$ \\
				$\lvert\beta _{0V}\rvert$                       &                  $\cdots$                   &          $\cdots$          & $(90.00 \pm 90.00)^\circ$  \\ \hline
			\end{tabular}
		\end{center}\label{Tab:data0VV}
	%
		\caption{The branching ratios for the $\chi_{c1} \to VV$ decays within  1$\sigma$ error (in units of $10^{-4}$). $^\sharp$indicates that experimental data was not used to derive the numerical results. }
		\centering
			\begin{center}
		\renewcommand\arraystretch{0.98}\tabcolsep 0.15in
		\begin{tabular}{lccc}
			\hline\hline
			                                                & Exp. data \cite{ParticleDataGroup:2024cfk} &          Symmetry          &     Including breaking     \\ \hline
			$\mathcal{B}(\chi_{c1}\to \rho^{+}\rho^{-})$    &                  $\cdots$                  &      $12.35\pm 1.92$       &      $15.18\pm 4.96$       \\
			$\mathcal{B}(\chi_{c1}\to \rho^{0}\rho^{0})$    &                  $\cdots$                  &       $6.17\pm 0.96$       &       $7.59\pm 2.48$       \\
			$\mathcal{B}(\chi_{c1}\to \rho\rho)$            &                  $\cdots$                  &      $18.52\pm 2.87$       &      $22.77\pm 7.43$       \\
			$\mathcal{B}(\chi_{c1}\to K^{*+}K^{*-})$        &                  $\cdots$                  &      $11.85\pm 1.84$       &      $13.22\pm 3.21$       \\
			$\mathcal{B}(\chi_{c1}\to  K^{*0}\bar{K}^{*0})$ &              $14.00\pm 4.00$               &      $11.84\pm 1.84$       &      $13.20\pm 3.20$       \\
			$\mathcal{B}(\chi_{c1}\to  K^{*}\bar{K}^{*})$   &                  $\cdots$                  &      $23.69\pm 3.68$       &      $26.42\pm 6.41$       \\
			$\mathcal{B}(\chi_{c1}\to \phi \phi)$           &               $4.26\pm 0.21$               &   $5.24\pm 0.73^\sharp$    &       $4.26\pm 0.21$       \\
			$\mathcal{B}(\chi_{c1}\to \phi \omega)$         &               $0.27\pm 0.04$               &       $0.27\pm 0.04$       &       $0.27\pm 0.04$       \\
			$\mathcal{B}(\chi_{c1}\to \omega \omega)$       &               $5.70\pm 0.70$               &       $5.70\pm 0.70$       &       $5.70\pm 0.70$       \\ \hline
			$a^{V}_{11}$ $(10^{-3})$                        &                  $\cdots$                  &       $6.96\pm 0.54$       &       $7.59\pm 1.17$       \\
			$a^{V}_{21}$ $(10^{-3})$                        &                  $\cdots$                  &       $1.06\pm 0.10$       &       $1.05\pm 0.19$       \\
			$b^{V}_{1}$ $(10^{-3})$                         &                  $\cdots$                  &          $\cdots$          &       $1.17\pm 1.14$       \\
			$\lvert\alpha _{1V}\rvert$                      &                  $\cdots$                  & $(105.42 \pm 28.65)^\circ$ & $(134.65 \pm 45.26)^\circ$ \\
			$\lvert\beta _{1V}\rvert$                       &                  $\cdots$                  &          $\cdots$          & $(47.56 \pm 47.56)^\circ$  \\ \hline
		\end{tabular}
			\end{center}\label{Tab:data1VV}
	\end{table}
	
  \begin{table}[t]
\caption{The branching ratios for the $\chi_{c2} \to VV$ decays  within  1$\sigma$ error (in units of $10^{-4}$). $^\sharp$indicates that experimental data was not used to derive the numerical results.
}
	\centering
	\begin{center}
	\renewcommand\arraystretch{1.2}\tabcolsep 0.15in
	\begin{tabular}{lccc}
		\hline\hline
		                                                & Exp. data \cite{ParticleDataGroup:2024cfk} &         Symmetry          &     Including breaking     \\ \hline
		$\mathcal{B}(\chi_{c2}\to \rho^{+}\rho^{-})$    &                  $\cdots$                  &      $27.08\pm 5.20$      &      $18.38\pm 8.04$       \\
		$\mathcal{B}(\chi_{c2}\to \rho^{0}\rho^{0})$    &                  $\cdots$                  &      $13.54\pm 2.60$      &       $9.19\pm 4.02$       \\
		$\mathcal{B}(\chi_{c2}\to \rho\rho)$            &                  $\cdots$                  &      $40.62\pm 7.80$      &      $27.57\pm 12.06$      \\
		$\mathcal{B}(\chi_{c2}\to K^{*+}K^{*-})$        &                  $\cdots$                  &      $26.04\pm 5.00$      &      $20.39\pm 7.37$       \\
		$\mathcal{B}(\chi_{c2}\to  K^{*0}\bar{K}^{*0})$ &              $22.00\pm 9.00$               &      $26.00\pm 4.99$      &      $20.36\pm 7.36$       \\
		$\mathcal{B}(\chi_{c2}\to K^{*}\bar{K}^{*})$    &                  $\cdots$                  &      $52.03\pm 9.99$      &      $40.76\pm 14.74$      \\
		$\mathcal{B}(\chi_{c2}\to \phi \phi)$           &              $12.30\pm 0.70$               &      $12.30\pm 0.70$      &      $12.30\pm 0.70$       \\
		$\mathcal{B}(\chi_{c2}\to \phi \omega)$         &              $0.097\pm 0.028$              &     $0.097\pm 0.028$      &      $0.097\pm 0.028$      \\
		$\mathcal{B}(\chi_{c2}\to \omega \omega)$       &               $8.60\pm 1.00$               &  $13.57\pm 2.36^\sharp$   &       $8.60\pm 1.00$       \\ \hline
		$a^{V}_{12}$ $(10^{-3})$                        &                  $\cdots$                  &      $16.31\pm 1.87$      &      $13.73\pm 3.09$       \\
		$a^{V}_{22}$ $(10^{-3})$                        &                  $\cdots$                  &      $0.98\pm 0.17$       &       $0.98\pm 0.48$       \\
		$b^{V}_{2}$ $(10^{-3})$                         &                  $\cdots$                  &         $\cdots$          &       $2.92\pm 2.59$       \\
		$\lvert\alpha _{2V}\rvert$                      &                  $\cdots$                  & $(90.00 \pm 90.00)^\circ$ & $(90.00 \pm 90.00)^\circ$  \\
		$\lvert\beta _{2V}\rvert$                       &                  $\cdots$                  &         $\cdots$          & $(131.21 \pm 48.70)^\circ$ \\ \hline
	\end{tabular}
			\end{center}\label{Tab:data2VV}
	\end{table}

The results of $\chi_{c0}\to VV$ with the SU(3) flavor symmetry  show good agreement with present experimental data  within $1\sigma$ error,  and their error ranges are small.
However, for $\chi_{c1}\to VV$  decays, the constrained $a^V_{11}$ from $\chi_{c1}\to  K^{*0}\bar{K}^{*0},\phi \omega,\omega \omega$ can not explain the data of $\chi_{c1}\to \phi \phi$ within $1\sigma$ error, nevertheless, one parameter $a^V_{11}$ could explain all four data if the error expand from $1\sigma$ to $1.1\sigma$ errors. As for $\chi_{c2}\to VV$  decays, all four experimental data can be explained by one parameter  $a^V_{22}$ within $1.5\sigma$ errors.
 The maximum values of  $\frac{a^{V}_{2J}}{a^{V}_{1J}}$ across all datasets reach 30\% for the  $\chi _{c0}$ decays, 17\% for the  $\chi _{c1}$ decays, and 8\% for the  $\chi _{c2}$ decays. They indicate  the dominant role of the $a^{V}_{1J}$ amplitudes in the $\chi _{cJ}\to VV$ decays. Despite some theoretical shortcomings, the SU(3) flavor symmetry   captures the essential correlations and still offers reasonable predictions.  Thus, the results based on the SU(3) flavor symmetry reflect universal trends, with a small uncertainty in the data distribution, offering useful reference values.

Now turn to  the  results including the SU(3) flavor breaking terms.  All present experimental data of the $\chi_{c0,1,2}\to VV$ decays can be explained within $1\sigma$ error bar if including  the SU(3) flavor breaking terms.  The maximum ratio of breaking contributions, i.e. $\frac{b^{V}_{J}}{a^{V}_{1J}}$, is 55\% for the $\chi _{c0}$ decays, 33\% for the $\chi _{c1}$ decays, and 48\% for the $\chi _{c2}$ decays. Compared to the $\chi _{cJ}\rightarrow  {P}{P}$ decays,  the allowed  breaking contributions for $\chi _{c0}$ and $\chi _{c2}$ are more significant in the $VV$ decays at present, and they notably alter the distribution of branching ratios. 
The fundamental reason is that there are five relevant non-perturbative parameters and only four experimental data in the $\chi_{c0/c1/c2} \to VV$ decays, and the  four experimental data are not enough to constrain on the five parameters. So $b^{V}_{J}$ and $\beta _{JV}$ are not well determined due to the lack of experimental values for the $\rho\rho$ channels, making it difficult to accurately predict the contribution of the breaking. And in the decays of $\chi _{c2}\to  {V}{V}$, the constraints on parameters are further weakened due to the large error of the $K^{*0}\bar{K}^{*0}$ channel, such as $a^{V}_{22}$ and $\alpha _{2V}$. Therefore, as the breaking contributions are included, the uncertainty in the predicted results also becomes larger, especially for some decay channels like $\chi_{c0,1,2}\to \rho\rho$ and $\chi_{c1,2}\to K^{*}\bar{K}^{*}$. However, the $\phi\phi$, $\phi\omega$, and $\omega\omega$ channels with good agreement still offer valuable insights, the predicted results reflect the basic distribution range, provide early useful reference points. Under same approach, the error ranges will decrease with more experimental inputs in the future. 

\begin{figure}[b]
\begin{center}
\includegraphics[scale=0.2]{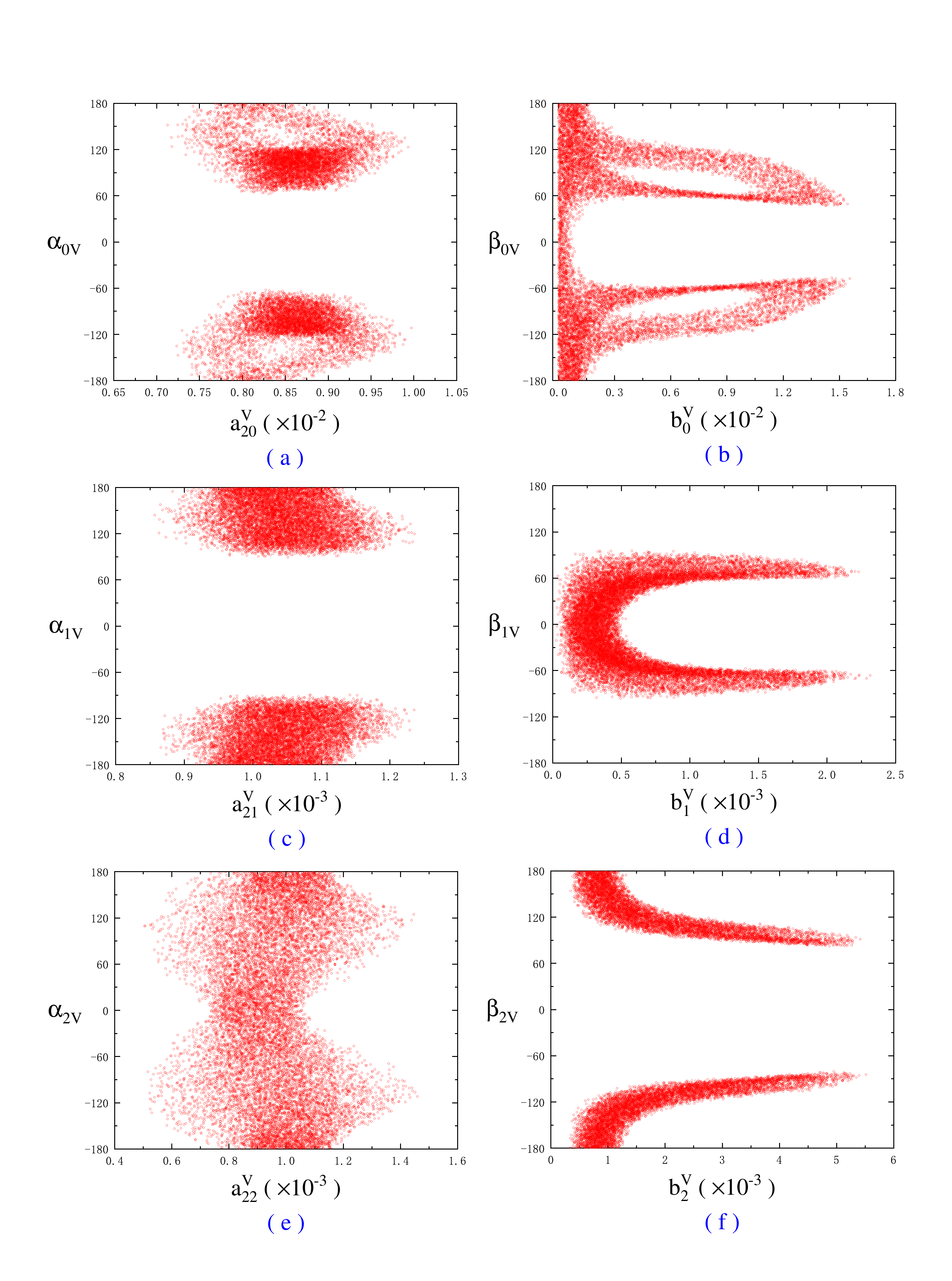}
\end{center}
\caption{The parameter correlations in the  $\chi _{c0,1,2}\rightarrow VV$ decays.
 }	\label{fig:VV}
\end{figure}

Fig. \ref{fig:VV}  shows the relationship between the  constrained symmetry/breaking parameters $(a^{V}_{2J},b^{V}_{J})$ and they their associated phases $(\alpha _{JV},\beta _{JV})$. One can see that, after satisfying all relevant experimental data, the allowed spaces are still large.  This may indicate that the  non-perturbative coefficients have not been very well limited by present experimental data within $1\sigma$ error, further leading to an increase in the uncertainty range of some predicted branching ratio results.
	
Now we discuss the non-yet-measured $\chi_{cJ}\to \rho\rho$ decays. Previous works  given that $\mathcal{B}(\chi_{c0}\to \rho\rho)=(1.88\pm1.80)\times 10^{-3}$ and $\mathcal{B}(\chi_{c2}\to \rho\rho)=(2.41\pm2.22)\times 10^{-3}$ in a general factorization scheme \cite{Zhao:2005im}, and    $\mathcal{B}(\chi_{c1}\to \rho\rho)$ lies in $[26,54]\times 10^{-4}$ by the intermediate meson loop contributions \cite{Liu:2009vv}.  Our predictions are consistent with  them,  and  our ones of the $\chi_{c1,2}\to \rho\rho$ decays are relatively more precise ranges.

\section{ The $\chi_{cJ} \to PV,PT$ Decays  } \label{sec:PVPT}
\subsection{Amplitude relations of the $\chi_{cJ} \to PV$ and $PT$ decays}

The decays of $\chi_{c0,2}\to PV$ and $PT$ are suppressed by  helicity selection rule, furthermore, the G-parity or isospin conservation exists in these decays. Rendering conventional theoretical approaches are significantly challenged in addressing these suppressed decay modes. Due to the  helicity selection rule, no many $\chi_{cJ}\to PV, PT$ decay channels have been measured.
The decay of $\chi_{c0}\to PV$ and $PT$ are vanish due to charmed vector current conservation and parity conservation forbidden by spin-parity conservation \cite{LHCb:2023evz}, and lack the robust phenomenological constraints in the absence of experimental observable values \cite{Ablikim:2006vm,LHCb:2023evz}. Therefore, only the $\chi_{c1}$ and $\chi_{c2}$ decays will be discussed in this section.

The SU(3) flavor symmetry/breaking approach is still used to obtain the branching ratio results of $\chi_{cJ}\to PV$ and $PT$.  The tensor meson octet states  can be written as
	\begin{eqnarray}
		T &=&\left(
		\begin{array}{ccc}
			\frac{a ^{0}_{2}}{\sqrt{2}}+\frac{f ^{8}_{2}}{\sqrt{6}}+\frac{f^{0}_{2}}{\sqrt{3%
			}} & a ^{+}_{2} & K^{*+}_{2} \\
			a ^{-}_{2} & -\frac{a ^{0}_{2}}{\sqrt{2}}+\frac{f ^{8}_{2}}{\sqrt{6}}+\frac{f^{0}_{2}}{\sqrt{3}} & K^{*0}_{2} \\
			K^{*-}_{2} & \bar{K}^{*0}_{2} & -\frac{2f^{8}_{2}}{\sqrt{6}}+\frac{f^{0}_{2}}{\sqrt{3}}%
		\end{array}
		\right). \label{T}
	\end{eqnarray}
The mixing of $f_{2}$-$f'_{2}$ is described as
\begin{eqnarray}
	f_{2}  &=&\left( f ^{8}_{2}\cos \theta _{T}-\allowbreak f^{0}_{2}\sin \theta
	_{T}\right),  \\
	f'_{2} &=&\left( f ^{8}_{2}\sin \theta _{T}+\allowbreak f^{0}_{2}\cos \theta _{T}\right),
\end{eqnarray}
and the results of $\theta _{T}=(27 \pm 1)^\circ$ is selected from PDG \cite{ParticleDataGroup:2024cfk}. Following the same theory as given in Sec. \ref{subsec:PPVVAR}, we derive the decay amplitudes for $PT$ and $PV$ channels,  and they are  listed in Tab. \ref{Tab:AmpPVT}.
	\begin{table}[h]
	\renewcommand\arraystretch{1.5}
	\tabcolsep 0.2in
	\caption{The SU(3) symmetry and breaking amplitudes of $\chi_{cJ}\to PV$ and $\chi_{cJ}\to PT$}
	\begin{center}
		\resizebox{\textwidth}{!}{
			\begin{tabular}{lll}
				\hline\hline
				Decay modes                                                      & Symmetry amplitudes                                                                       & Breaking amplitudes                                                                  \\ \hline
				$\chi _{cJ}\rightarrow  \pi^{+}\rho^{-}/\pi^{+}a^{-}_{2}$        & $2a^{MM}_{1J}$                                                                            & $2b^{MM}_{J}$                                                                        \\
				$\chi _{cJ}\rightarrow  \pi^{-}\rho^{+}/\pi^{-}a^{+}_{2}$        & $2a^{MM}_{1J}$                                                                             & $2b^{MM}_{J}$                                                                        \\
				$\chi _{cJ}\rightarrow  K^{+}K^{*-}/K^{-}K^{*+}_{2}$             & $2a^{MM}_{1J}$                                                                            & $-b^{MM}_{J}$                                                                        \\
				$\chi _{cJ}\rightarrow  K^{-}K^{*+}/K^{-}K^{*+}_{2}$             & $2a^{MM}_{1J}$                                                                             & $-b^{MM}_{J}$                                                                        \\
				$\chi _{cJ}\rightarrow  K^{0}\bar{K}^{*0}/K^{0}\bar{K}^{*0}_{2}$ & $2a^{MM}_{1J}$                                                                             & $-b^{MM}_{J}$                                                                        \\
				$\chi _{cJ}\rightarrow  \bar{K}^{0}K^{*0}/\bar{K}^{0}K^{*0}_{2}$ & $2a^{MM}_{1J}$                                                                             & $-b^{MM}_{J}$                                                                        \\
				$\chi _{cJ}\rightarrow  \pi^{0}\rho^{0}/\pi^{0}a^{0}_{2}$        & $2a^{MM}_{1J}$                                                                             & $2b^{MM}_{J}$                                                                        \\
				$\chi _{cJ}\rightarrow  \eta_{1} \omega_{1}/\eta_{1} f^{0}_{2}$  & $2a^{MM}_{1J}+6a^{MM}_{2J}$                                                               & $0$                                                                                  \\
				$\chi _{cJ}\rightarrow  \eta_{1} \omega_{8}/\eta_{1} f^{8}_{2}$  & $0$                                                                                       & $2\sqrt{2}b^{MM}_{J}$                                                                \\
				$\chi _{cJ}\rightarrow  \eta_{8} \omega_{1}/\eta_{8} f^{0}_{2}$  & $0$                                                                                       & $2\sqrt{2}b^{MM}_{J}$                                                                \\
				$\chi _{cJ}\rightarrow  \eta_{8} \omega_{8}/\eta_{8} f^{8}_{2}$  & $2a^{MM}_{1J}$                                                                            & $-2b^{MM}_{J}$                                                                       \\ \hline
				$\chi _{cJ}\rightarrow  \eta \phi/\eta f'_{2} $               & $2a^{MM}_{1J}(\cos(\theta_{P}-\theta_{V/T}))+6a^{MM}_{2J}\sin\theta_{P}\sin\theta_{V/T}$  & $-2b^{MM}_{J}(\cos\theta_{P}\cos\theta_{V/T}+\sqrt{2}\sin(\theta_{P}+\theta_{V/T}))$ \\
				$\chi _{cJ}\rightarrow  \eta \omega/\eta f_{2} $                 & $2a^{MM}_{1J}(-\sin(\theta_{P}-\theta_{V/T}))-6a^{MM}_{2J}\sin\theta_{P}\cos\theta_{V/T}$ & $-2b^{MM}_{J}(\cos\theta_{P}\sin\theta_{V/T}-\sqrt{2}\cos(\theta_{P}+\theta_{V/T}))$ \\
				$\chi _{cJ}\rightarrow  \eta' \phi/\eta' f'_{2} $          & $2a^{MM}_{1J}(\sin(\theta_{P}-\theta_{V/T}))-6a^{MM}_{2J}\cos\theta_{P}\sin\theta_{V/T}$  & $-2b^{MM}_{J}(\sin\theta_{P}\cos\theta_{V/T}-\sqrt{2}\cos(\theta_{P}+\theta_{V/T}))$ \\
				$\chi _{cJ}\rightarrow  \eta' \omega/\eta' f_{2} $            & $2a^{MM}_{1J}(\cos(\theta_{P}-\theta_{V/T}))+6a^{MM}_{2J}\cos\theta_{P}\cos\theta_{V/T}$  & $-2b^{MM}_{J}(\sin\theta_{P}\sin\theta_{V/T}-\sqrt{2}\sin(\theta_{P}+\theta_{V/T}))$ \\ \hline
			\end{tabular}}
	\end{center}\label{Tab:AmpPVT}
\end{table}

 \subsection{Numerical results of the $\chi_{cJ}\to PV$ decays }  \label{subsec:PV}
 The neutral $PV$ channels of the $\chi_{c1}$ and $\chi_{c2}$ decays, such as $\pi^{0}\rho^{0}$ and $\eta\omega$, etc, are forbidden by $C$-parity conservation. Furthermore, the $\chi_{c2}\to PV$ decays are not only suppressed by helicity selection rule, but also suffer from G-parity or isospin/$U$-spin conservation \cite{Liu:2009vv}.   Although these processes lack experimental observations, we still provide phenomenological predictions under existing models. It is worth mentioning that many $PV$ and $PT$ processes are difficult to effectively constrain theoretically, so we only provide numerical predictions based on relevant experimental  data from PDG \cite{ParticleDataGroup:2024cfk}. Moreover, the experimental limits are significantly less than the theoretical non-perturbative parameters that need to be determined, and effective results cannot be obtained without additional constraints. So in the following contents, we will refer to the parameter ratios of $PP$ and $VV$ processes to limit the upper bounds of parameters, in order to obtain early prediction results.

For the $\chi_{c1}\to PV$ decays, only two modes of $\chi_{c1}\to K\bar{K}^*$   have been measured \cite{ParticleDataGroup:2024cfk}. Their experimental data are listed in the   second column of Tab. \ref{Tab:data1PV}. In these decays, relevant non-perturbative coefficients $a^{PV}_{21}$ and $b^{PV}_{1}$  lack effective experimental constraints.  Referring  to our results of $\chi_{cJ}\to PP,VV$,  the constraints of $\frac{a^{PV}_{21}}{a^{PV}_{11}}\leq 45\%$ and  $\frac{b^{PV}_{1}}{a^{PV}_{11}}\leq 55\%$ are  imposed to obtain the predictions. The branching ratio predictions results of $\chi_{c1}\to PV$ via the SU(3) flavor symmetry are shown in the third column of Tab. \ref{Tab:data1PV}, while ones including the breaking results are presented in the last column of Tab. \ref{Tab:data1PV}.   From Tab. \ref{Tab:data1PV}, one can see that  many errors of our predictions  are  quite large, since there are no other experimental data to constrain $a^{PV}_{21}$, $b^{PV}_{1}$ and corresponding two phases.
\begin{table}[ht]
 \caption{The branching ratios for the $\chi_{c1} \to PV$ decays within  1$\sigma$ error (in units of $10^{-3}$).}
 \centering
 \begin{center}
 \renewcommand\arraystretch{0.98}\tabcolsep 0.15in
 \begin{tabular}{lccc}
 	\hline\hline
 	                                                   & Exp. data  \cite{ParticleDataGroup:2024cfk} &         Symmetry          &    Including breaking     \\ \hline
 	$\mathcal{B}(\chi_{c1}\to  \pi^{\pm}\rho^{\mp})$   &                  $\cdots$                   &      $1.12\pm 0.11$       &      $2.78\pm 2.65$       \\
 	$\mathcal{B}(\chi_{c1}\to \pi^{0}\rho^{0})$        &                  $\cdots$                   &      $0.56\pm 0.05$       &      $1.39\pm 1.32$       \\
 	$\mathcal{B}(\chi_{c1}\to K^{\pm}K^{*\mp})$        &               $1.21\pm 0.23$                &      $1.08\pm 0.10$       &      $1.08\pm 0.10$       \\
 	$\mathcal{B}(\chi_{c1}\to K^{0}\bar{K}^{*0}+c.c.)$ &               $1.03\pm 0.15$                &      $1.08\pm 0.10$       &      $1.08\pm 0.10$       \\
 	$\mathcal{B}(\chi_{c1}\to \eta \phi$)              &                  $\cdots$                   &      $0.21\pm 0.17$       &      $0.40\pm 0.40$       \\
 	$\mathcal{B}(\chi_{c1}\to \eta \omega)$            &                  $\cdots$                   &      $0.47\pm 0.34$       &      $1.16\pm 1.15$       \\
 	$\mathcal{B}(\chi_{c1}\to \eta' \phi )$            &                  $\cdots$                   &      $0.64\pm 0.64$       &      $0.87\pm 0.87$       \\
 	$\mathcal{B}(\chi_{c1}\to \eta' \omega )$          &                  $\cdots$                   &      $0.77\pm 0.77$       &      $1.93\pm 1.93$       \\ \hline
 	$a^{PV}_{11}$ $(10^{-3})$                          &                  $\cdots$                   &      $4.68\pm 0.32$       &      $5.12\pm 1.65$       \\
 	$a^{PV}_{21}$ $(10^{-3})$                          &                  $\cdots$                   &      $1.11\pm 1.11$       &      $1.46\pm 1.46$       \\
 	$b^{PV}_{1}$ $(10^{-3})$                           &                  $\cdots$                   &         $\cdots$          &      $1.85\pm 1.85$       \\
 	$\lvert\alpha _{1PV}\rvert$                        &                  $\cdots$                   & $(90.00 \pm 90.00)^\circ$ & $(90.00 \pm 90.00)^\circ$ \\
 	$\lvert\beta _{1PV}\rvert$                         &                  $\cdots$                   &         $\cdots$          & $(90.00 \pm 90.00)^\circ$ \\ \hline
 \end{tabular}
\end{center}\label{Tab:data1PV}
%
 \caption{The branching ratios for the $\chi_{c2} \to PV$ decays  within  1$\sigma$ error (in units of $10^{-4}$). $^\sharp$indicates that experimental data was not used to derive the numerical results.}
 \centering
 \begin{center}
 	\renewcommand\arraystretch{0.98}\tabcolsep 0.15in
 	\begin{tabular}{lccc}
 		\hline\hline
 		                                                   & Exp. data  \cite{ParticleDataGroup:2024cfk} &         Symmetry          &     Including breaking     \\ \hline
 		$\mathcal{B}(\chi_{c2}\to  \pi^{\pm}\rho^{\mp})$   &               $0.06\pm 0.04$                &   $1.45\pm 0.15^\sharp$   &       $0.06\pm 0.04$       \\
 		$\mathcal{B}(\chi_{c2}\to \pi^{0}\rho^{0})$        &                  $\cdots$                   &      $0.73\pm 0.08$       &       $0.03\pm 0.02$       \\
 		$\mathcal{B}(\chi_{c2}\to K^{\pm}K^{*\mp})$        &               $1.46\pm 0.21$                &      $1.40\pm 0.15$       &       $1.40\pm 0.15$       \\
 		$\mathcal{B}(\chi_{c2}\to K^{0}\bar{K}^{*0}+c.c.)$ &               $1.27\pm 0.27$                &      $1.40\pm 0.15$       &       $1.40\pm 0.15$       \\
 		$\mathcal{B}(\chi_{c2}\to \eta \phi)$              &                  $\cdots$                   &      $0.28\pm 0.23$       &       $0.91\pm 0.48$       \\
 		$\mathcal{B}(\chi_{c2}\to \eta \omega)$            &                  $\cdots$                   &      $0.59\pm 0.43$       &       $0.09\pm 0.09$       \\
 		$\mathcal{B}(\chi_{c2}\to \eta' \phi )$            &                  $\cdots$                   &      $0.81\pm 0.81$       &       $1.55\pm 1.25$       \\
 		$\mathcal{B}(\chi_{c2}\to \eta' \omega )$          &                  $\cdots$                   &      $1.01\pm 1.01$       &       $0.31\pm 0.31$       \\ \hline
 		$a^{PV}_{12}$ $(10^{-3})$                          &                  $\cdots$                   &      $2.60\pm 0.19$       &       $1.91\pm 0.18$       \\
 		$a^{PV}_{22}$ $(10^{-3})$                          &                  $\cdots$                   &      $0.61\pm 0.61$       &       $0.46\pm 0.46$       \\
 		$b^{PV}_{2}$ $(10^{-3})$                           &                  $\cdots$                   &         $\cdots$          &       $1.38\pm 0.21$       \\
 		$\lvert\alpha _{2PV}\rvert$                        &                  $\cdots$                   & $(90.00 \pm 90.00)^\circ$ & $(90.00 \pm 90.00)^\circ$  \\
 		$\lvert\beta _{2PV}\rvert$                         &                  $\cdots$                   &         $\cdots$          & $(169.60 \pm 10.31)^\circ$ \\ \hline
 	\end{tabular}
 \end{center}\label{Tab:data2PV}
 \end{table}

For the $\chi_{c2}\to PV$ decays, three decay modes,  $\chi_{c2}\to K^{\pm}K^{*\mp}, K^{0}\bar{K}^{*0}+c.c.,\pi^{\pm}\rho^{\mp}$ have been measured \cite{ParticleDataGroup:2024cfk}, and  they are listed in the   second column of Tab. \ref{Tab:data2PV}.
As given in the third column of \ref{Tab:data2PV}, within the SU(3) flavor symmetry,  and the constrained $a_{12}^{PV}$ from $\mathcal{B}(\chi_{c2}\to K^{\pm}K^{*\mp}, K^{0}\bar{K}^{*0}+c.c.)$ can not explain the experimental data of $\mathcal{B}(\chi_{c2}\to  \pi^{\pm}\rho^{\mp})$. Furthermore, with the SU(3) flavor symmetry,  $\mathcal{B}(\chi_{c2}\to  \pi^{\pm}\rho^{\mp})$ is on the order of $\mathcal{O}(10^{-4})$, and $\mathcal{B}(\chi_{c2}\to \pi^{0}\rho^{0})$ is on the order of $\mathcal{O}(10^{-5})$. If considering the SU(3) flavor breaking, their allowed minimum values are down by an order of magnitude. This means that a large breaking effect is needed to explain the experimental data of $\mathcal{B}(\chi_{c2}\to  \pi^{\pm}\rho^{\mp})$.

If considering the SU(3) flavor breaking  and setting $\frac{a^{PV}_{22}}{a^{PV}_{12}}\leq 45\%$ as well as $\frac{b^{PV}_{2}}{a^{PV}_{12}}\leq 80\%$,  all  three experimental data are explained at the same time.  Noted that three  experimental data can not be explained together when $\frac{b^{PV}_{2}}{a^{PV}_{12}}\leq 63\%$.
The predictions including the breaking contributions are presented in the last column of Tab. \ref{Tab:data2PV}.  One can see that  $a^{PV}_{12}$, $b^{PV}_{2}$ and  $\beta _{2PV}$ are well constrained by three experimental data within $1\sigma$ error. Nevertheless, there is no any constraint on  $a^{PV}_{22}$ and $\alpha _{2PV}$, which due to large error in some $\chi_{c2}\to PV$ decays.  Some predicted results fail to exhibit the anticipated theoretical suppression, as this effect is obscured by substantial theoretically uncertainties.  And the branching ratios of $\chi_{c2}\to \pi\rho$ and $\chi_{c2}\to \eta \omega$ are significantly smaller compared to other decay channels when the flavor breaking effects are included. If we do not consider the theoretically forbidden scenario, this may be a signal that is suppressed by helicity selection rule or other theories.

Moreover, please noted that the predicted branching ratios of $K^{\pm}K^{*\mp}$ and $K^{0}\bar{K}^{*0}+c.c.$ are completely consistent in both symmetry and breaking cases.  The main reason is that their shared amplitude structure $2a_{1J}^{PV}-b^{PV}_Je^{i\beta_J^{PV}}$  and   the effective constraints on $\big|2a_{1J}^{PV}-b^{PV}_Je^{i\beta_J^{PV}}\big|$ are actually obtained from the data of $K^{\pm}K^{*\mp}$ and $K^{0}\bar{K}^{*0}+c.c.$ decays at present.  The similar situation appear in the $\chi_{c1,2} \to PT$ decays.


 \subsection{Numerical results of the  $\chi_{cJ}\to PT$ decays }  \label{subsec:PT}
   	For the $PT$ decays, there are no $C/G$-parity conservation imposes prohibitive constraints, and helicity selection rule does not cause significant suppression. This theoretical landscape enables richer predictions for viable decay channels. Experimental measurements for $\chi _{c1}\to  PT$ and  $\chi _{c2}\to  PT$ decays are compiled in the second columns of Tab. \ref{Tab:data1PT} and Tab. \ref{Tab:data2PT}, respectively. SU(3) symmetry predictions occupy the third columns, and including  breaking cases presented in the final columns.
 \begin{table}[htb]
    	\caption{The branching ratios for the $\chi_{c1} \to PT$ decays within  1$\sigma$ error (in units of $10^{-3}$).}
    	\centering
    	\begin{center}
    		\renewcommand\arraystretch{0.98}\tabcolsep 0.15in
    		\begin{tabular}{lccc}
    			\hline\hline
    			                                                       & Exp. data  \cite{ParticleDataGroup:2024cfk} &         Symmetry          &    Including breaking     \\ \hline
    			$\mathcal{B}(\chi_{c1}\to \pi^{\pm}a^{\mp}_{2})$       &                  $\cdots$                   &      $1.42\pm 0.04$       &      $3.03\pm 2.14$       \\
    			$\mathcal{B}(\chi_{c1}\to \pi^{0}a^{0}_{2})$           &                  $\cdots$                   &      $0.71\pm 0.02$       &      $1.51\pm 1.07$       \\
    			$\mathcal{B}(\chi_{c1}\to K^{\pm}K^{*\mp}_{2})$        &               $1.61\pm 0.31$                &      $1.34\pm 0.04$       &      $1.34\pm 0.04$       \\
    			$\mathcal{B}(\chi_{c1}\to K^{0}\bar{K}^{*0}_{2}+c.c.)$ &               $1.17\pm 0.20$                &      $1.33\pm 0.04$       &      $1.33\pm 0.04$       \\
    			$\mathcal{B}(\chi_{c1}\to \eta f'_{2})$                &                  $\cdots$                   &      $0.23\pm 0.03$       &      $0.42\pm 0.31$       \\
    			$\mathcal{B}(\chi_{c1}\to \eta f_{2})$                 &               $0.67\pm 0.11$                &      $0.61\pm 0.05$       &      $0.67\pm 0.11$       \\
    			$\mathcal{B}(\chi_{c1}\to \eta' f'_{2} )$              &                  $\cdots$                   &      $0.60\pm 0.07$       &      $0.76\pm 0.76$       \\
    			$\mathcal{B}(\chi_{c1}\to \eta' f_{2} )$               &                  $\cdots$                   &      $0.99\pm 0.21$       &      $1.49\pm 1.49$       \\ \hline
    			$a^{PT}_{11}$ $(10^{-3})$                              &                  $\cdots$                   &      $5.54\pm 0.20$       &      $6.09\pm 1.16$       \\
    			$a^{PT}_{21}$ $(10^{-3})$                              &                  $\cdots$                   &      $1.18\pm 0.25$       &      $1.60\pm 1.60$       \\
    			$b^{PT}_{1}$ $(10^{-3})$                               &                  $\cdots$                   &         $\cdots$          &      $1.92\pm 1.92$       \\
    			$\lvert\alpha _{1PT}\rvert$                            &                  $\cdots$                   & $(25.78 \pm 25.78)^\circ$ & $(90.00 \pm 90.00)^\circ$ \\
    			$\lvert\beta _{1PT}\rvert$                             &                  $\cdots$                   &         $\cdots$          & $(90.00 \pm 90.00)^\circ$ \\ \hline
    		\end{tabular}
    	\end{center}\label{Tab:data1PT}
%
    	\caption{The branching ratios for the $\chi_{c2} \to PT$ decays  within  1$\sigma$ error (in units of $10^{-3}$). $^\sharp$indicates that experimental data was not used to derive the numerical results.}
    	\centering
    	\begin{center}
    		\renewcommand\arraystretch{0.98}\tabcolsep 0.15in
    		\begin{tabular}{lccc}
    			\hline\hline
    			                                                       & Exp. data \cite{ParticleDataGroup:2024cfk} &         Symmetry          &    Including breaking     \\ \hline
    			$\mathcal{B}(\chi_{c2}\to \pi^{\pm}a^{\mp}_{2})$       &               $1.80\pm 0.60$               &      $1.50\pm 0.03$       &      $2.16\pm 0.24$       \\
    			$\mathcal{B}(\chi_{c2}\to \pi^{0}a^{0}_{2})$           &               $1.31\pm 0.35$               &   $0.75\pm 0.02^\sharp$   &      $1.08\pm 0.12$       \\
    			$\mathcal{B}(\chi_{c2}\to K^{\pm}K^{*\mp}_{2})$        &               $1.51\pm 0.13$               &      $1.41\pm 0.03$       &      $1.41\pm 0.03$       \\
    			$\mathcal{B}(\chi_{c2}\to K^{0}\bar{K}^{*0}_{2}+c.c.)$ &               $1.27\pm 0.17$               &      $1.41\pm 0.03$       &      $1.41\pm 0.03$       \\
    			$\mathcal{B}(\chi_{c2}\to \eta f'_{2})$                &                  $\cdots$                  &      $0.37\pm 0.15$       &      $0.45\pm 0.36$       \\
    			$\mathcal{B}(\chi_{c2}\to \eta f_{2})$                 &                  $\cdots$                  &      $0.41\pm 0.25$       &      $0.77\pm 0.62$       \\
    			$\mathcal{B}(\chi_{c2}\to \eta' f'_{2})$               &                  $\cdots$                  &      $0.37\pm 0.32$       &      $0.81\pm 0.81$       \\
    			$\mathcal{B}(\chi_{c2}\to \eta' f_{2} )$               &                  $\cdots$                  &      $0.70\pm 0.70$       &      $1.59\pm 1.59$       \\ \hline
    			$a^{PT}_{12}$ $(10^{-3})$                              &                  $\cdots$                  &      $8.77\pm 0.30$       &      $9.10\pm 0.70$       \\
    			$a^{PT}_{22}$ $(10^{-3})$                              &                  $\cdots$                  &      $1.13\pm 1.13$       &      $2.19\pm 2.19$       \\
    			$b^{PT}_{2}$ $(10^{-3})$                               &                  $\cdots$                  &         $\cdots$          &      $2.86\pm 2.14$       \\
    			$\lvert\alpha _{2PT}\rvert$                            &                  $\cdots$                  & $(90.00 \pm 90.00)^\circ$ & $(90.00 \pm 90.00)^\circ$ \\
    			$\lvert\beta _{2PT}\rvert$                             &                  $\cdots$                  &         $\cdots$          & $(43.54 \pm 43.54)^\circ$ \\ \hline
    		\end{tabular}
    	\end{center}\label{Tab:data2PT}
    \end{table}

The $\eta f'_{2}$, $\eta f_{2}$, $\eta' f'_{2}$ and $\eta' f_{2}$ channels exhibit sensitivity to the $a^{PT}_{2J}$ amplitude parameter. However, the absence of experimental constraints make it difficult to determine the accuracy of $a^{PT}_{2J}$ and $b^{PT}_{J}$ in the simulation, especially there is no experimental data in $\chi_{c2}\to PT$ that can limit the range of $a^{PT}_{22}$. To address this, we impose the upper limits, $\frac{a^{PT}_{2J}}{a^{PT}_{1J}}\leq 45\%$ and $\frac{b^{PT}_{J}}{a^{PT}_{1J}}\leq 55\%$,  based on analysis of $PP$ and $VV$ channels.

For the $\chi_{c1}\to PT$ decays, the $\eta f_{2}$ channel provides critical constraints on $a^{PT}_{21}$ and $\alpha _{1PT}$, improving the accuracy of predicting $\eta f'_{2}$, $\eta' f'_{2}$ and $\eta' f_{2}$ channels in the SU(3) flavor symmetry case. After considering the SU(3) flavor breaking, $a^{PT}_{21}$ and  $b^{PT}_{1}$ are not constrained well, and corresponding phase  $\alpha _{1PT}$ and $\beta _{1PT}$ are completely unrestricted.  These bring about large errors in $\mathcal{B}(\chi_{c1}\to \pi^{\pm}a^{\mp}_{2})$, 	$\mathcal{B}(\chi_{c1}\to \pi^{0}a^{0}_{2})$, $\mathcal{B}(\chi_{c1}\to \eta' f'_{2} )$ and  $\mathcal{B}(\chi_{c1}\to \eta' f_{2} )$, and the measurements of these branching ratios could further constrained relevant  non-perturbative coefficients.

For the $\chi_{c2}\to PT$ decays, in the SU(3) flavor symmetry case,  the experiential data of $\mathcal{B}(\chi_{c2}\to \pi^{\pm}a^{\mp}_{2})$, $\mathcal{B}(\chi_{c2}\to K^{\pm}K^{*\mp}_{2})$ and $\mathcal{B}(\chi_{c2}\to K^{0}\bar{K}^{*0}_{2}+c.c.)$  give quite strong constraint on $a^{PT}_{12}$, and  the data of $\mathcal{B}(\chi_{c2}\to \pi^{0}a^{0}_{2})$ can not be explained with other three ones together. Since there is no experimental constraint on $a^{PT}_{22}$ and $\alpha _{2PT}$,   the sources of errors are increased and there are  relatively large ranges of uncertainty in the $\chi_{c2}\to \eta f'_{2}$, $\eta f_{2}$, $\eta' f'_{2}$ and $\eta' f_{2}$ channels.

After considering the SU(3) flavor breaking, all four experimental data of the $\chi_{c2}\to PT$ decays can be explained together within $1\sigma$ error bar. And the current experimental data  impose some restrictions on $b^{PT}_{2}$ and $\beta _{2PT}$, but they are not strong. In fact, for all $PV$ and $PT$ processes, effective limitations have only been imposed on $a^{PV,PT}_{1J}$ within the existing experimental constraints, which at least ensures that all predicted results reflect the basic characteristics of the decay channels.


\section{Conclusion} \label{sec:Conclusion}
Previous measurements by BESIII were based on the accumulated 448 million $\psi(3686)$ decays \cite{BESIII:2020nme}, which allowed access to $\chi _{cJ}$ decays through radiative decays $\psi(3686) \to  \gamma\chi_{cJ}$. With a current data sample of 2.7 billion $\psi(3686)$ events collected by the BESIII detector  \cite{BESIII:2024lks}, a more detailed analysis of two body $\chi_{cJ}$ decays is now possible. This provides an opportunity to test the SU(3) symmetry and gain deeper insights into their decay mechanisms.

In this work, we have investigated the  $\chi _{c0,2}\to {P}{P}$, $\chi _{c0,1,2}\to {V}{V}$ and $\chi _{c1,2}\to PV,PT$ decays with the SU(3) flavor symmetry/breaking approach.  We have obtained the  amplitude relations   in both SU(3) flavor symmetry case and further including SU(3) flavor breaking case. Because the advantages of symmetry approach, inconclusive intermediate decay mechanisms can be avoided from discussion. All branching ratios, including the predicted results for not-yet-measured or not-well-measured channels, have been presented in the work.

Our study shows that the SU(3) symmetry approach work well in the two body decays of charmonium states $\chi _{cJ}$ at present, which are typically difficult to compute using traditional QCD methods.
  Specifically, the $\chi_{cJ}\rightarrow {\eta}{\eta'}$ and $\chi_{cJ}\rightarrow {\omega}{\phi}$ are doubly OZI-suppressed, resulting in branching ratios for these decays  are obviously smaller than those for the singly OZI-suppressed decays $\chi_{cJ}\rightarrow {\eta}{\eta}$, $\eta\eta'$, $\omega\omega$ and $\phi\phi$ \cite{BESIII:2017rqn,Zheng:2017ffd,BESII:2011hcd,Achasov:2024eua}. Our results are consistent with the suppression of the $\eta\eta'$ and $\omega\phi$ channels in the contributions of $a^{M}_{2J}$ and align well with experimental data for these channels.

Our predictions not only contribute to future experimental measurements, but also offer valuable reference points for future theoretical works that aims to refine these calculations.
Furthermore, with the high luminosity of the current experiments \cite{Charm-TauFactory:2013cnj,Peng:2020orp,BESIII:2024lbn}, we can expect a continuous stream of interesting experimental results to emerge, including precise measurements of branching ratios and potentially new insights into the decay mechanisms of $\chi_{cJ}$ decaying into meson pairs.
	
\section*{ACKNOWLEDGEMENTS}
The work was supported by the
National Natural Science Foundation of China (Nos. 12175088, 12305100 and 12365014) and Graduate Innovation Fund Project of Jiangxi Provincial Department of Education (No. YJS2024091).

\section*{References}
\renewcommand{\baselinestretch}{1.5}

\end{document}